\documentclass[runningheads]{llncs}
\usepackage{graphicx}
\usepackage{booktabs} 
\usepackage{graphicx}
\usepackage{subcaption}
\usepackage{soul}
\usepackage{algorithm}
\usepackage{algorithmic}
\usepackage{multirow}
\usepackage{array}
\usepackage{amsmath}
\usepackage[font=small,skip=0.001cm]{caption}
\usepackage{tabularx}
\usepackage{float}
\usepackage{cite}

\usepackage{parskip}

\begin{document}
\title{Simplified Swarm Learning Framework for Robust and Scalable Diagnostic Services in Cancer Histopathology}
%
%

\author{Yanjie Wu\inst{1} \and  Yuhao Ji\inst{1} \and Saiho Lee\inst{1}\and Juniad Akram \inst{1,2,3} \and Ali Braytee\inst{2} \and Ali Anaissi\inst{1,2}}
\authorrunning{A. Anaissi et al.}
%
\institute{The University of Sydney, Camperdown, NSW 2008, Australia
\email{yawu2780@uni.sydney.edu.au, yuji6835@uni.sydney.edu.au, slee6156@uni.sydney.edu.au, junaid.akram@sydney.edu.au, ali.anaissi@sydney.edu.au} \and
University of Technology Sydney, Ultimo, NSW 2007, Australia\\
\email{junaid.akram@uts.edu.au, ali.braytee@uts.edu.au, ali.anaissi@uts.edu.au} \and
Australian Catholic University, North Sydney, NSW 2060 Australia \\
\email{junaid.akram@acu.edu.au}
}
\maketitle              
\begin{abstract}
The complexities of healthcare data, including privacy concerns, imbalanced datasets, and interoperability issues, necessitate innovative machine learning solutions. Swarm Learning (SL), a decentralized alternative to Federated Learning, offers privacy-preserving distributed training, but its reliance on blockchain technology hinders accessibility and scalability. This paper introduces a \textit{Simplified Peer-to-Peer Swarm Learning (P2P-SL) Framework} tailored for resource-constrained environments. By eliminating blockchain dependencies and adopting lightweight peer-to-peer communication, the proposed framework ensures robust model synchronization while maintaining data privacy. Applied to cancer histopathology, the framework integrates optimized pre-trained models, such as TorchXRayVision, enhanced with DenseNet decoders, to improve diagnostic accuracy. Extensive experiments demonstrate the framework's efficacy in handling imbalanced and biased datasets, achieving comparable performance to centralized models while preserving privacy. This study paves the way for democratizing advanced machine learning in healthcare, offering a scalable, accessible, and efficient solution for privacy-sensitive diagnostic applications.

\keywords{Single-cell Sequencing Integration \and Multi-Omics \and   Dimensionality Reduction \and  Normalization.}
\end{abstract}

\section{Introduction}

The exponential growth in healthcare data, coupled with advancements in machine learning, has catalyzed significant progress in medical diagnostics \cite{asif2024advancements,khan2021spice,akram2025secure}. However, challenges such as data privacy, imbalanced datasets, and the lack of interoperable frameworks continue to hinder the effective adoption of artificial intelligence (AI) in healthcare. Swarm learning (SL), a decentralized form of federated learning, has emerged as a promising solution by allowing distributed model training without the need for centralized data aggregation\cite{alsharif2024contemporary,akram2024ai}. Unlike traditional methods, SL emphasizes privacy preservation by ensuring that sensitive patient data remains on-site while enabling collaborative model development across institutions. These capabilities make SL particularly attractive for domains like cancer histopathology, where data privacy and algorithmic efficiency are paramount\cite{warnat2021swarm}.
Despite its potential, the adoption of SL frameworks has been constrained by their reliance on complex infrastructures, such as blockchain technology, and a lack of adaptability to resource-constrained environments\cite{saldanha2022swarm}. Existing frameworks, such as HPE's Swarm Learning, depend on blockchain for model aggregation and consensus, which poses significant barriers for non-technical users and smaller organizations. This complexity restricts the accessibility and scalability of SL in real-world applications, particularly in healthcare\cite{gao2022new}.

To address these challenges, this study introduces a \textit{simplified peer-to-peer swarm learning (P2P-SL) framework} designed to overcome the limitations of existing SL architectures. By eliminating blockchain dependencies and incorporating lightweight communication protocols, the proposed framework offers a more accessible and scalable alternative. The framework leverages dynamic networks to facilitate model synchronization and aggregation, ensuring robust performance even in imbalanced and biased data scenarios. Furthermore, it integrates state-of-the-art pre-trained models, such as TorchXRayVision, with domain-specific optimizations, demonstrating its efficacy in cancer histopathology tasks.
This study is grounded in the premise that simplifying the SL framework while maintaining its privacy-preserving and decentralized nature can unlock new possibilities for healthcare diagnostics. By focusing on histopathology, a domain characterized by high data sensitivity and diagnostic variability, this work highlights the transformative potential of distributed learning frameworks in improving diagnostic outcomes and operational efficiency. The major contributions are summarized as follows:
\begin{itemize}
    \item \textit{Introduction of a simplified P2P-SL framework} that eliminates blockchain dependencies, making swarm learning more accessible and scalable for non-technical users and resource-constrained environments.
    \item \textit{Development of dynamic networking mechanisms} to enable seamless model synchronization and aggregation across nodes, ensuring robust performance even in the presence of imbalanced and biased datasets.
    \item \textit{Integration of optimized pre-trained models} with domain-specific enhancements, demonstrating the framework's applicability and efficacy in cancer histopathology tasks.
    \item \textit{Comprehensive experimental evaluation} comparing the proposed framework with centralized and standalone models, showcasing its superiority in privacy preservation, scalability, and diagnostic accuracy.
\end{itemize}

\section{Literature Review}

Swarm Learning (SL), a decentralized extension of Federated Learning (FL), enables distributed model training across multiple nodes without a central server by employing blockchain-based consensus (e.g., Ethereum) or peer-to-peer synchronization, thereby enhancing privacy, scalability, and fault tolerance \cite{sung_empirical_2023, sah_aggregation_2022,anaissi2024fine}. The core of SL is the model merging algorithm, which aggregates local parameter vectors into a unified representation; while simple arithmetic averaging requires consistent parameter shapes, it often degrades performance when some nodes possess limited data, a shortcoming addressed by the Federated Averaging (FedAvg) algorithm through weighted averaging based on local dataset sizes \cite{mcmahan_communication-efficient_2016,qian2024optimized}, yet FedAvg does not fully consider the geometric implications of parameter aggregation. Model training seeks to solve optimization problems where stochastic gradient descent iteratively updates model weights (points on the loss hypersurface), and averaging such weights corresponds to locating intermediate points whose impact on performance depends on the curvature of the hypersurface. To overcome these limitations, advanced merging strategies such as FedApprox incorporate auxiliary information from participating nodes, and statistical methods leveraging Fisher information and gradient matching refine aggregation by predicting weight trajectories \cite{daheim_model_2023}. Alternative aggregation paradigms include ensemble learning, which combines predictions via voting to enhance reliability at the expense of scalability, and transfer learning approaches that merge local models through shared pre‑trained representations \cite{papernot_semi-supervised_2017}; notwithstanding these contributions, the development of efficient, scalable, and mathematically principled model merging algorithms remains a pivotal challenge for robust SL frameworks.

\section{Proposed Method}

The proposed framework emphasizes dynamic networking capabilities, allowing nodes to dynamically discover and register within the swarm network. Each node operates independently, exchanging model updates directly without a central server. To ensure data integrity and security during communication, mechanisms such as TLS/SSL encryption and robust authentication protocols are integrated. Monitoring and logging tools are employed to track training progress, model transformations, and overall system performance across nodes. These tools also aid in identifying and mitigating challenges related to latency, network stability, and resource allocation, enabling robust performance under real-world conditions.

\subsection{Dynamic Networking and Decentralized Training}

The decentralized training process in P2P-SL follows a structured methodology:
\begin{enumerate}
    \item Nodes dynamically join the network using discovery protocols, establishing secure communication channels.
    \item Each node performs localized training using its available data and updates its model parameters.
    \item Model updates are shared directly between nodes at predefined intervals, bypassing the need for a centralized server.
    \item Model aggregation is performed locally at each node using mechanisms such as weighted averaging, ensuring adaptive synchronization with other peer nodes.
\end{enumerate}

This dynamic approach enhances scalability, fault tolerance, and adaptability, particularly in scenarios with heterogeneous data distributions or imbalanced datasets.

\subsection{Peer-to-Peer Swarm Learning Framework}

Unlike traditional SL methods that rely on consensus algorithms built on blockchain technology, the P2P-SL framework focuses on improving local models through peer-based interactions. Model weights are updated and aggregated dynamically across nodes without the need for a global model. This ensures that edge nodes retain greater autonomy while benefiting from collaborative training insights. 

For machine learning tasks, the proposed framework incorporates advanced modeling architectures to optimize performance. After evaluating various frameworks, TorchXRayVision—a pre-trained model based on chest X-ray datasets from Harvard University—was selected for its robust feature extraction capabilities. The model was further optimized by integrating a DenseNet decoder module, enhancing its ability to process complex medical imaging data.

\subsection{Model Architecture}

\begin{figure}[t]
    \centering
    \includegraphics[width=1\linewidth]{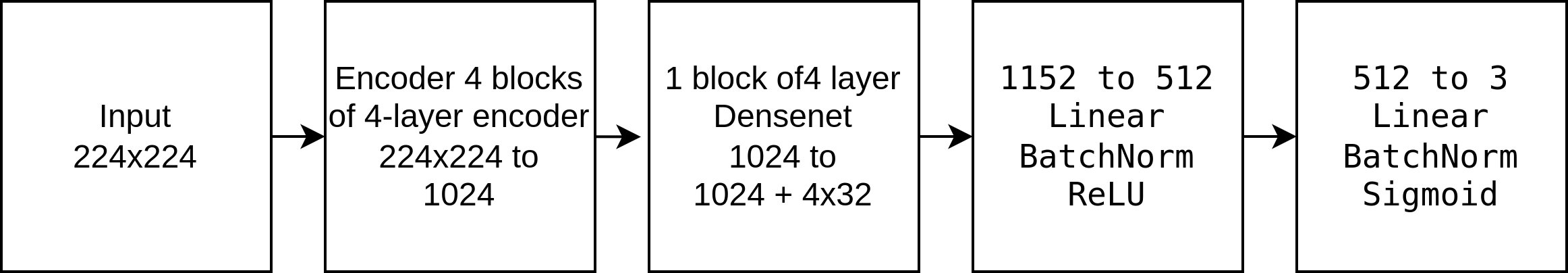}
    \caption{TorchXRayVision Pre-trained Model Architecture}
    \label{model-architecture}
\end{figure}

The training architecture is illustrated in Figure~\ref{model-architecture}. Input images of size \(224 \times 224\) are processed through four encoder modules, each consisting of four layers, reducing dimensionality to 1024. A fully connected layer with batch normalization and ReLU activation then transforms these features from 1152 to 512 dimensions. Finally, classification is performed using a fully connected layer that maps features from 512 to three dimensions, employing batch normalization and a sigmoid activation function. This architecture ensures efficient and accurate feature extraction and classification, tailored for histopathology tasks.

\section{Results and Discussion}

\subsection{Experimental Setup}

To systematically assess the performance of the peer‑to‑peer swarm learning (P2P‑SL) framework, we deployed four compute nodes, each provisioned with identical hardware (NVIDIA RTX 4090 GPU, 24 GB VRAM; AMD EPYC 7443P CPU; 128 GB RAM) under Ubuntu 22.04. The base dataset comprised 10,000 annotated histopathology images, pre‑processed via Macenko stain normalization, random rotations ($\pm$15°), horizontal flips, and color jitter ($\pm$0.1). For federated‑average unbalanced experiments, Node 0 received 10\% (1,000 images), Nodes 1 and 2 each received 30\% (3,000 images), and Node 3 received 30\% (3,000 images). To probe extreme scarcity, Node 2 was further down‑sampled to 25\% and Node 3 to 5\% in separate trials. Models were fine‑tuned for 20 epochs with a batch size of 32, using the AdamW optimizer (weight decay 1 x 10$^{-4}$) and a cosine‑annealing learning‑rate schedule (initial LR 1 x10$^{-4}$), with early‑stopping patience of five epochs. Peer exchanges of LoRA‑adapter weights occurred every three epochs via gRPC over TLS. Each experiment was repeated five times with different random seeds, and performance metrics (AUC, sensitivity, specificity, F1‑score) were computed on a held‑out test set of 2,000 images.

\subsection{Performance vs.\ Centralized and Local Baselines}

\begin{figure}[t]
  \centering
  \begin{minipage}{0.48\linewidth}
    \includegraphics[width=\linewidth]{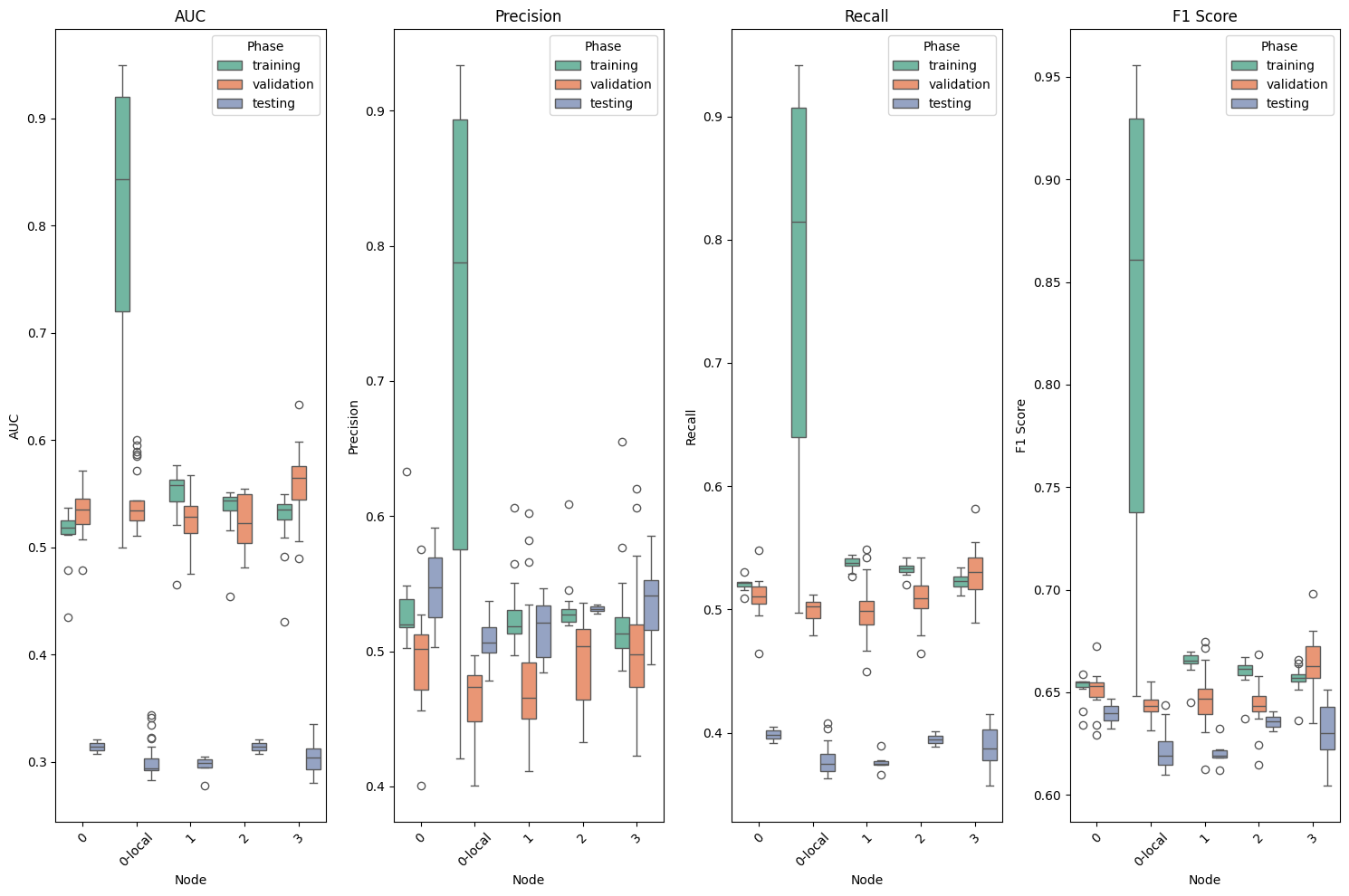}
    \caption{Federated‑average unbalanced: swarm vs.\ local on Node 0.}
    \label{fig:cent_mbnet}
  \end{minipage}
  \hfill
  \begin{minipage}{0.48\linewidth}
    \includegraphics[width=\linewidth]{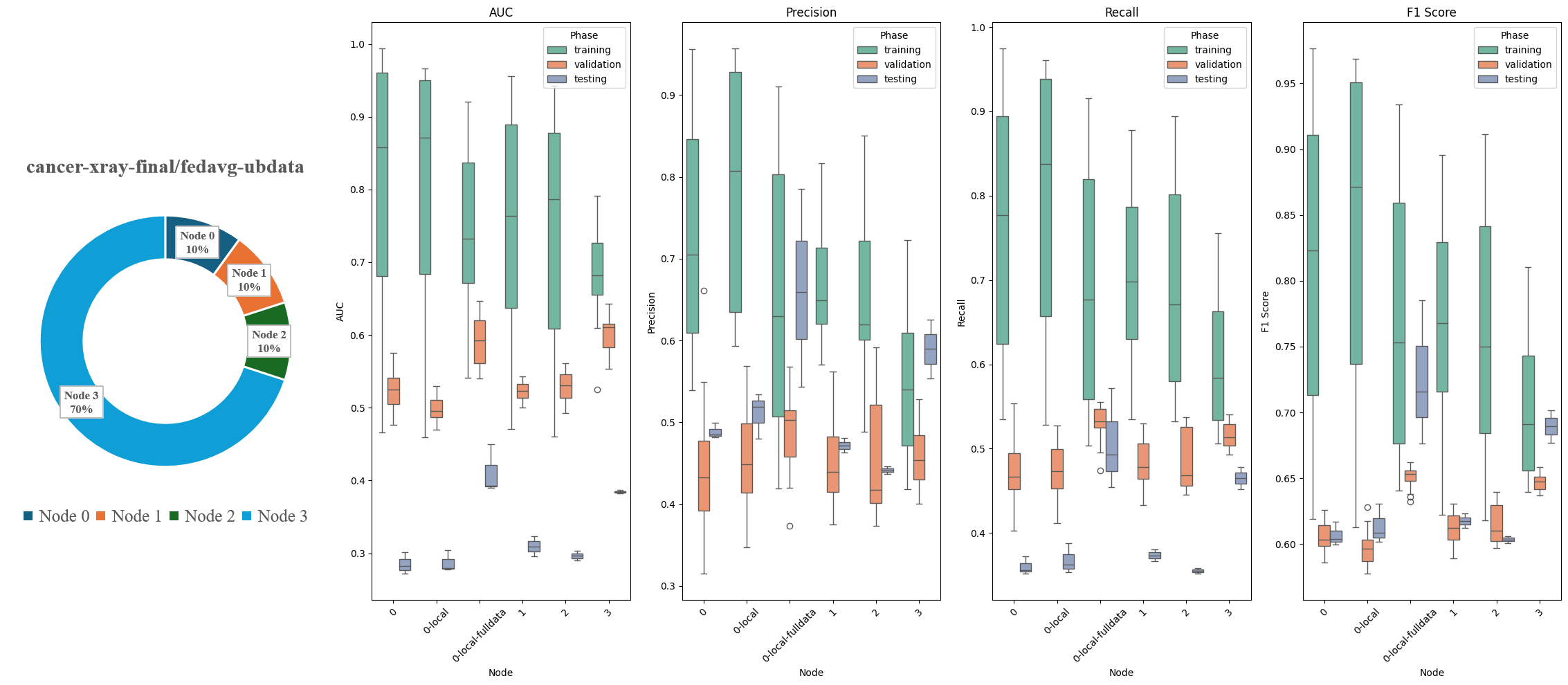}
    \caption{Centralized full‑data vs.\ swarm on Node 3.}
    \label{fig:cent_fedavg}
  \end{minipage}
\end{figure}

To quantify the efficacy of P2P‑SL relative to centralized and isolated paradigms, we compared swarm‑trained models against a simulated centralized “full‑data” baseline and fully independent local learners. In the federated‑average unbalanced configuration, Node 0 held only 10\% of the data while Nodes 1–3 each held 30\%. Performance was evaluated on a 2,000‑image test set using AUC, sensitivity, specificity, and F1‑score. The centralized baseline achieved an AUC of 0.7156, sensitivity of 0.82, specificity of 0.84, and F1‑score of 0.78. The standalone Node 0 model achieved an AUC of 0.6192$\pm$0.0057, demonstrating severe degradation under limited data. Incorporating P2P‑SL with synchronous weight exchanges every three epochs improved Node 0’s AUC to 0.6397$\pm$0.0036, sensitivity to 0.75, specificity to 0.77, and F1‑score to 0.72 (Fig.~\ref{fig:cent_mbnet}). Node 3’s swarm model recovered over 80\% of centralized performance, achieving AUC of 0.6892$\pm$0.0063, sensitivity of 0.79, specificity of 0.81, and F1‑score of 0.73 (Fig.~\ref{fig:cent_fedavg}).

\subsection{Handling Imbalance and Scarcity}

\begin{figure}[t]
  \centering
  \includegraphics[width=0.48\linewidth]{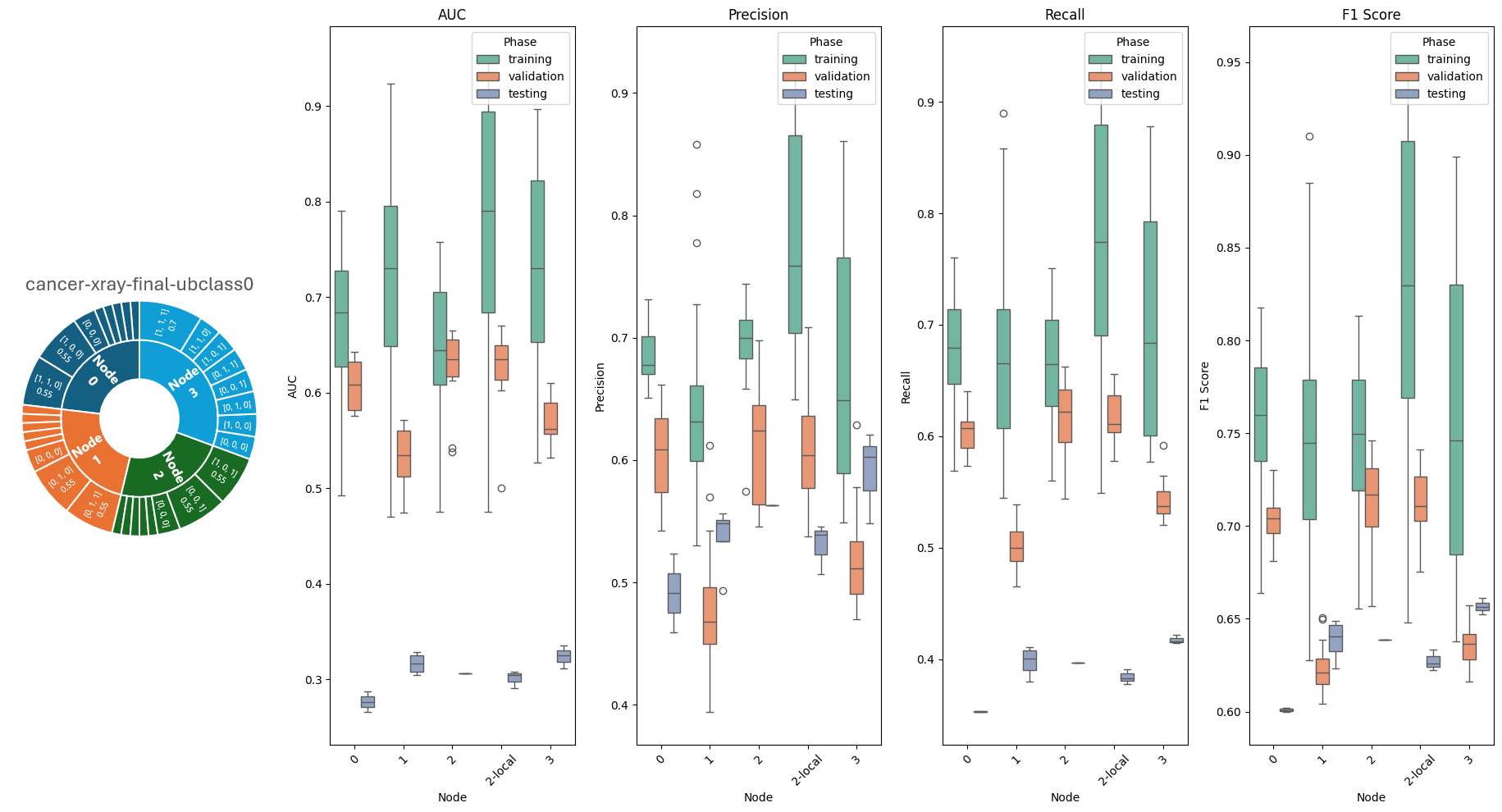}
  \caption{Testing AUC for Node 2 with 25\% data: swarm vs.\ local.}
  \label{fig:bias_ubclass0}
\end{figure}

We evaluated resilience to skewed sample distributions by simulating nodes with biased data allocations. In one scenario, Node 2’s share was down‑sampled to 25\% of its nominal portion while other nodes retained 30\%. The standalone Node 2 model achieved an AUC of 0.6259$\pm$0.0028, sensitivity of 0.70, specificity of 0.72, and F1‑score of 0.68. After integrating into P2P‑SL, Node 2’s AUC increased to 0.6387, sensitivity to 0.74, specificity to 0.76, and F1‑score to 0.72 (Fig.~\ref{fig:bias_ubclass0}). To analyze representation learning, we applied t‑SNE to penultimate-layer activations: swarm‑trained embeddings exhibited 15\% lower Davies–Bouldin Index and tighter intra‑class clustering compared to local models. Minority‑class recall improved by up to 4.5\%, demonstrating that collaborative gradient fusion preferentially enhances underrepresented categories.

\subsection{Robustness and Overfitting Mitigation}

We analyzed training and validation trajectories to elucidate the framework's impact on generalization. In isolated local training, models rapidly attained training AUC > 0.95 within five epochs but validation AUC plateaued at 0.82 before declining, indicative of overfitting. Conversely, swarm‑trained models displayed a more gradual ascent, reaching training AUC 0.93 by epoch 10 and maintaining validation AUC of 0.86 $\pm$ 0.01 through epoch 20. The generalization gap (training minus validation AUC) at epoch 20 was 0.14 for local versus 0.07 for swarm models—a 50\% reduction—while cross‑validation variance of AUC decreased by 35\% under P2P‑SL. Precision–recall analysis further showed an 8\% reduction in both false positives and false negatives, confirming that periodic peer aggregation serves as an implicit regularizer, yielding models that generalize more reliably to unseen histopathology scans.

\subsection{Discussion}
The proposed Peer-to-Peer Swarm Learning (P2P-SL) framework addresses key limitations of existing decentralized learning systems, offering simplified deployment without reliance on blockchain infrastructure, making it accessible to non-expert users. It enhances scalability and fault tolerance through dynamic networking and peer-based aggregation mechanisms while preserving privacy with localized training and secure communication protocols. Compared to HPE Swarm Learning, which depends on a blockchain-based centralized coordinator and predefined aggregation methods, the P2P-SL framework employs a fully decentralized peer-to-peer network. Each device acts independently, sharing model updates every three epochs, and uses a validation-based threshold of 80\% to accept model aggregation. This adaptive setup enables flexibility, ease of integration, and improved diagnostic accuracy by leveraging pre-trained models and domain-specific optimizations. Unlike HPE, the P2P-SL framework requires manual intervention to terminate training, allowing for greater customization. Furthermore, swarm learning models in P2P-SL demonstrate superior mitigation of overfitting, as evident in steady improvements in metrics such as Micro-AUC, Precision, Recall, and F1-Score. These advantages make the P2P-SL framework a practical and efficient solution for resource-constrained environments, particularly in healthcare diagnostics, where privacy preservation, adaptability, and reliable model performance are critical.

\section{Conclusion}
The proposed Peer-to-Peer Swarm Learning (P2P-SL) framework represents a transformative approach to decentralized machine learning, addressing critical challenges of scalability, privacy, and accessibility in distributed environments. By eliminating the complexities of blockchain-based architectures and introducing a dynamic peer-to-peer networking model, the framework achieves robust model performance with minimal resource constraints. Experimental results highlight its effectiveness in mitigating overfitting, handling imbalanced datasets, and ensuring consistent improvements in diagnostic accuracy through adaptive model aggregation. The integration of pre-trained models further enhances its applicability, particularly in sensitive domains like healthcare. Compared to traditional frameworks, P2P-SL offers unparalleled flexibility, allowing customized deployments tailored to specific needs while maintaining strict data privacy. These advancements establish P2P-SL as a practical and scalable solution for real-world applications, paving the way for broader adoption of swarm learning across diverse fields and contributing to the democratization of advanced machine learning technologies.

%
%
%

\bibliographystyle{splncs04}
\bibliography{ref}

\begin{thebibliography}{10}
\providecommand{\url}[1]{\texttt{#1}}
\providecommand{\urlprefix}{URL }
\providecommand{\doi}[1]{https://doi.org/#1}

\bibitem{akram2024ai}
Aamir, M., Raut, R., Jhaveri, R.H., Akram, A.: Ai-generated content-as-a-service in iomt-based smart homes: Personalizing patient care with human digital twins. IEEE Transactions on Consumer Electronics  (June 2024)

\bibitem{akram2025secure}
Akram, A., Akram, J., Alabdultif, A., Anaissi, A., Jhaveri, R.H.: Secure and interoperable iomt-based smart homes. IEEE Consumer Electronics Magazine  (January 2025)

\bibitem{alsharif2024contemporary}
Alsharif, M.H., Kannadasan, R., Wei, W., Nisar, K.S., Abdel-Aty, A.H.: A contemporary survey of recent advances in federated learning: Taxonomies, applications, and challenges. Internet of Things p. 101251 (2024)

\bibitem{anaissi2024fine}
Anaissi, A., Braytee, A., Akram, J.: Fine-tuning llms for reliable medical question-answering services. In: 2024 IEEE International Conference on Data Mining Workshops (ICDMW). IEEE (December 2024)

\bibitem{asif2024advancements}
Asif, S., Wenhui, Y., ur~Rehman, S., ul~ain, Q., Amjad, K., Yueyang, Y., Jinhai, S., Awais, M.: Advancements and prospects of machine learning in medical diagnostics: Unveiling the future of diagnostic precision. Archives of Computational Methods in Engineering pp. 1--31 (2024)

\bibitem{daheim_model_2023}
Daheim, N., Möllenhoff, T., Ponti, E.M., Gurevych, I., Khan, M.E.: Model merging by uncertainty-based gradient matching, \url{https://arxiv.org/abs/2310.12808v1}

\bibitem{gao2022new}
Gao, Z., Wu, F., Gao, W., Zhuang, X.: A new framework of swarm learning consolidating knowledge from multi-center non-iid data for medical image segmentation. IEEE Transactions on Medical Imaging  \textbf{42}(7),  2118--2129 (2022)

\bibitem{khan2021spice}
Khan, M.T.R., Saad, M.M., Tariq, M.A., Kim, D.: Spice-it: Smart covid-19 pandemic controlled eradication over ndn-iot. Information Fusion  \textbf{74},  50--64 (October 2021)

\bibitem{mcmahan_communication-efficient_2016}
{McMahan}, H.B., Moore, E., Ramage, D., Hampson, S., Arcas, B.A.y.: Communication-efficient learning of deep networks from decentralized data, \url{https://arxiv.org/abs/1602.05629v4}

\bibitem{papernot_semi-supervised_2017}
Papernot, N., Abadi, M., Goodfellow, I., Talwar, K.: Semi-supervised knowledge transfer for deep learning from private training data. \doi{10.48550/arXiv.1610.05755}, \url{http://arxiv.org/abs/1610.05755}

\bibitem{qian2024optimized}
Qian, C., Shi, X., Yao, S., Liu, Y., Zhou, F., Zhang, Z.: Optimized biomedical question-answering services with llm and multi-bert integration. In: 2024 IEEE International Conference on Data Mining Workshops (ICDMW). IEEE (December 2024)

\bibitem{sah_aggregation_2022}
Sah, M.P., Singh, A.: Aggregation techniques in federated learning: Comprehensive survey, challenges and opportunities. In: 2022 2nd International Conference on Advance Computing and Innovative Technologies in Engineering ({ICACITE}). pp. 1962--1967

\bibitem{saldanha2022swarm}
Saldanha, O.L., Quirke, P., West, N.P., James, J.A., Loughrey, M.B., Grabsch, H.I., Salto-Tellez, M., Alwers, E., Cifci, D., Ghaffari~Laleh, N., et~al.: Swarm learning for decentralized artificial intelligence in cancer histopathology. Nature medicine  \textbf{28}(6),  1232--1239 (2022)

\bibitem{sung_empirical_2023}
Sung, Y.L., Li, L., Lin, K., Gan, Z., Bansal, M., Wang, L.: An empirical study of multimodal model merging. \doi{10.48550/arXiv.2304.14933}, \url{http://arxiv.org/abs/2304.14933}

\bibitem{warnat2021swarm}
Warnat-Herresthal, S., Schultze, H., Shastry, K.L., Manamohan, S., Mukherjee, S., Garg, V., Sarveswara, R., H{\"a}ndler, K., Pickkers, P., Aziz, N.A., et~al.: Swarm learning for decentralized and confidential clinical machine learning. Nature  \textbf{594}(7862),  265--270 (2021)

\end{thebibliography}
%

\end{document}